\documentclass{article}
\usepackage{spconfa4,amsmath,graphicx}

\usepackage{cite}
\usepackage{amssymb,amsfonts}
\usepackage{algorithmic}
\usepackage{textcomp}
\usepackage{xcolor}
\usepackage[T1]{fontenc}
\usepackage{url}
\usepackage{subcaption}
\usepackage{orcidlink}
\usepackage{siunitx}
\usepackage{verbatim}
\usepackage{booktabs}
\usepackage{multirow}
\usepackage{balance}

\usepackage{xurl}

\makeatletter
\renewcommand\footnotesize{%
  \@setfontsize\footnotesize{9}{10}%
}
\makeatother

\title{\uppercase{Training DeepFilterNet with Accurate Room Acoustic Simulations Improves Single-Channel Speech Enhancement}}
\name{Alessia Milo, Georg Götz, Steinar Gu$\eth{}$jónsson, Daniel Gert Nielsen, Jesper Pedersen, Finnur Pind }
\address{Treble Technologies\\
Reykjavík\\
Iceland}

\begin{document}
\ninept
\maketitle
%

\begin{abstract}

We investigate how the realism of synthetic room impulse response (RIR) datasets affects the training of DeepFilterNet3 for single-channel speech enhancement. We compare a DNS4 image-source-method (ISM) RIR dataset with a higher-acoustic-fidelity dataset generated using hybrid wave-based and geometrical acoustics simulation. Rather than isolating individual simulation factors, we compare complete RIR generation pipelines while keeping the enhancement model unchanged. Models are evaluated on unseen measured RIRs using objective speech enhancement metrics and downstream automatic speech recognition (ASR). Training with the higher-fidelity dataset consistently yields modest improvements in objective metrics and substantially lower ASR word error rates than the ISM dataset. Although the experiments do not attribute these gains to individual modelling components, they show that increasing the overall realism of synthetic acoustic training data improves the generalization of DeepFilterNet3 to unseen measured environments.

\end{abstract}

\begin{keywords}
Speech enhancement, Room impulse response (RIR) simulation, DeepFilterNet, Reverberation modeling, Automatic speech recognition (ASR)
\end{keywords}
\section{Introduction}
\label{sec:intro}

Single-microphone audio capture remains prevalent in consumer and embedded devices, where cost, size, and power constraints limit the use of arrays. As a result, robust speech enhancement under adverse acoustic conditions is essential, yet fundamentally challenging due to the lack of spatial information.

Recent advances in deep learning, including CRNN-based models \cite{tan2018convolutional} and large-scale training on datasets such as DNS \cite{dubey2022icassp2022deepnoise}, have significantly improved single-channel enhancement. However, these approaches typically rely on simplified RIRs generated via the image-source method \cite{Ko2017DataAugmentationReverberantSpeechForSpeechRecognition,Kim2017GenerationLargeScaleSimulatedUtterancesFarFieldASR,Maciejewski2020WHAMR}, which do not fully capture real acoustic conditions.

Efficient architectures such as DeepFilterNet \cite{schroter2022deepfilternet,schroeter2023deepfilternet3} enable real-time joint denoising and dereverberation and have shown strong performance across challenging scenarios \cite{wang2018supervised,wu2025egonoise,rosenbaum2024deepfiltering,westhausen2020dtln}. Nevertheless, the impact of RIR fidelity on enhancement performance remains underexplored.

Prior work has shown that increasing the realism of synthetic room acoustics can improve speech-processing performance \cite{bezzam2020study,arakawa2024quantifying,Srivastava2023HowToVirtuallyTrainYourSpeakerLocalizer,guso2025mb,Tang2022GWALargeDatasetForAudioProcessing}. Gusó et al.~\cite{guso2025mb} isolated factors such as frequency-dependent absorption and source directivity for DeepFilterNet3, whereas Aralikatti et al.~\cite{aralikatti2022synthetic} studied hybrid wave-geometric simulation for speech dereverberation.

Here, we compare two complete RIR generation pipelines: conventional image-source simulation and higher-fidelity hybrid simulation. The DeepFilterNet3 architecture and training procedure remain unchanged. Rather than isolating individual acoustic factors, we assess whether greater overall simulation realism improves generalization in objective enhancement metrics and downstream ASR on unseen measured RIRs. The results motivate future controlled studies of geometry, material modelling, frequency-dependent absorption, and wave-based simulation.

Our experiments evaluate both objective speech enhancement metrics and downstream automatic speech recognition (ASR) performance on unseen measured RIRs. While the present study does not attribute the observed improvements to any individual simulation factor, it demonstrates that training with a more realistic synthetic RIR dataset consistently improves generalization across all evaluated configurations. These findings complement previous studies on individual simulation factors and motivate future work aimed at disentangling the respective contributions of room geometry, material modelling, frequency-dependent absorption, and hybrid wave-based simulation.

\section{Acoustic simulation}
\label{sec:AcousticSimulation}

In data augmentation for speech enhancement, the choice of acoustic simulation method directly affects the spatial and spectral properties of the generated RIRs. Room-acoustic simulation approaches are commonly divided into geometrical acoustics (GA) and wave-based methods, with hybrid techniques combining both. GA methods, such as the image-source method (ISM) and ray tracing, approximate sound propagation using specular reflections and energy transport assumptions that are most valid at higher frequencies \cite{Savioja2015OverviewGA,Krokstad1968CalculatingTheAcousticalRoomResponseByTheUseOfRayTracing,Krokstad1983Fifteenyearsexperiencewithcomputerizedraytracing,Allen1979ImageMethodForEfficientlySimulatingSmallRoomAcoustics,Borish1984ExtensionOfTheImageModelToArbitraryPolyhedra}. However, they do not inherently capture wave phenomena such as diffraction, interference, or room modes, and diffraction in particular remains challenging to model accurately within GA frameworks \cite{Torres2001ComputationEdgeDiffractionRoomAcousticSimulations,Schissler2014HighOrderDiffractionAndDiffuseReflectionsForInteractiveSoundPropagation}. These limitations contribute to known discrepancies and variability across simulation tools \cite{Vorlaender2013ComputerSimulationsRoomAcoustics,Brinkmann2019RoundRobinOnRoomAcousticalSimulationAndAuralization}.

Wave-based methods instead solve the acoustic wave equation directly, enabling accurate modeling of diffraction, modal behavior, and frequency-dependent boundary conditions, which are especially relevant at low and mid frequencies \cite{Hamilton2016FiniteDifferenceVolumeWaveBasedRoomAcousticsThesis,Pind2019SpectralElementRoomAcousticSimulation,Prinn2023ReviewFEMRoomAcoustics,BottelDooren1995FDTDSimulationRoomAcoustic}. Although traditionally more computationally demanding, recent advances have made large-scale simulations increasingly feasible \cite{Wang2019DiscontinuousGalerkinRoomAcoustics,Pind2020TimeDomainSimExtendedReactingPorousDGFEM,Melander2024MassivelyParallelGalerkinRoomAcoustics}. Hybrid approaches combine both paradigms by applying wave-based solvers below a crossover frequency and GA above it, capturing low-frequency wave effects while retaining efficiency at higher frequencies \cite{Siltanen2010RaysOrWavesStrenghtsAndWeaknesses}. In data augmentation, this allows improved modeling of modal behavior, diffraction, and frequency-dependent decay while maintaining broadband coverage.

\section{Experiments}

\label{sec:experiment}

We compare an ISM RIR dataset with a higher-fidelity hybrid dataset and describe the corresponding DeepFilterNet3 training and evaluation protocol.

\subsection{ISM: Image-source dataset}

The simulated RIR dataset used in this work is based on the 48 kHz OpenSLR release (SLR28) and corresponds to the RIR data provided in the DNS Challenge 4 (DNS4) training set. The dataset comprises \num{60000} RIRs organized into three room-size categories (small, medium, and large), each containing 200 shoebox rooms with 100 impulse responses per room. In this dataset, the reverberation time used as a target varies from \SIrange{0.05}{1.0}{\second} and is used to calculate a single uniform wall absorption coefficient. 

\begin{figure}[]
\centering
\includegraphics[width=\columnwidth]{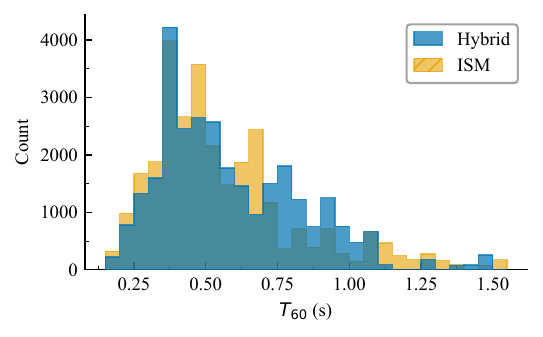}
\caption{Histogram of Sabine RT60 values for Hybrid and ISM matched datasets}
\label{fig:rt60}
\end{figure}

To match dataset size and reverberation-time distribution, we paired Hybrid and ISM rooms using Sabine RT$_{60}$ estimates derived from their metadata and selected the closest ISM matches. Figure~\ref{fig:rt60} shows the resulting distributions.

\subsection{Hybrid: High-fidelity dataset}

\begin{figure}[]
\centering
\includegraphics[width=\columnwidth]{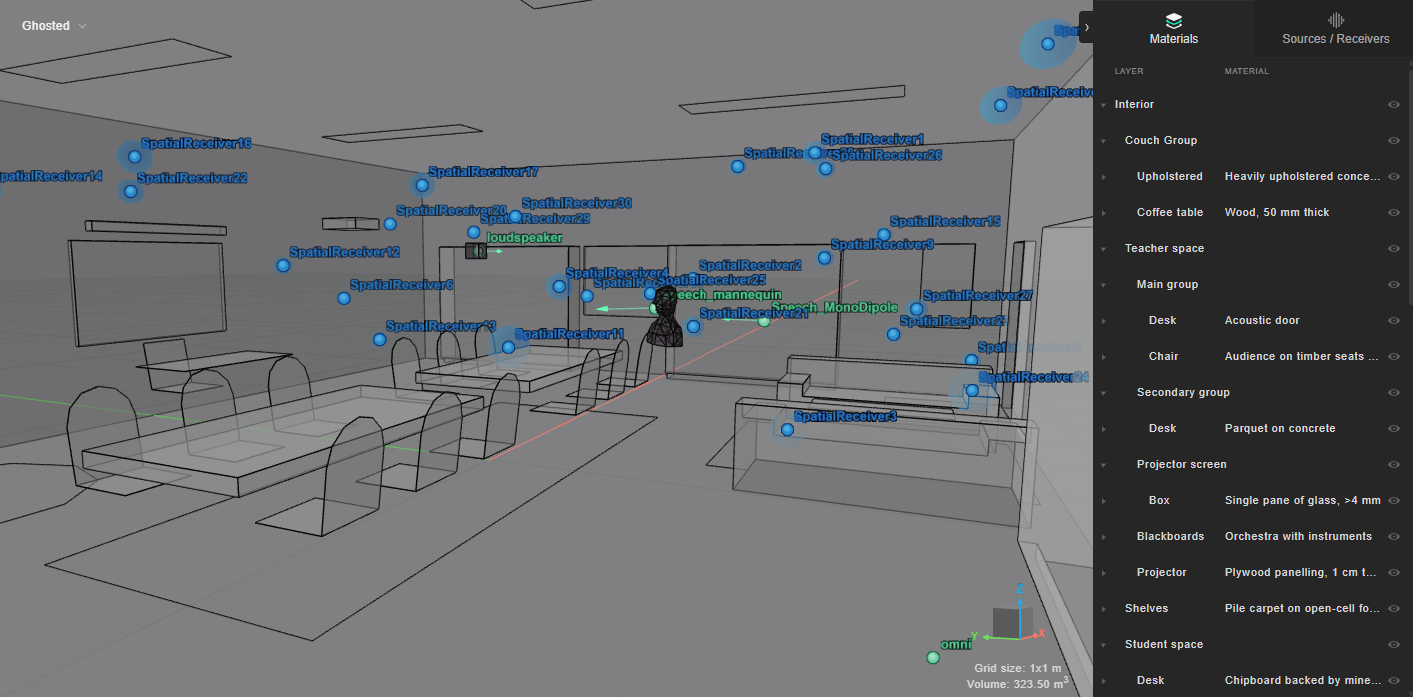}
\caption{Level of detail for one of the rooms simulated in the Hybrid dataset}
\label{fig:room}
\end{figure}

The hybrid dataset was generated using the Treble SDK, with all geometries and boundary-condition materials drawn from its built-in libraries. It comprises three subsets corresponding to different room types and size ranges: 133 living rooms (40–\SI{180}{\cubic\meter}), 103 classrooms (90–\SI{400}{\cubic\meter}), and 88 restaurants (300–\SI{1600}{\cubic\meter}). All rooms are furnished and exhibit diverse geometries and structural complexity, as shown in Fig. \ref{fig:room}.

Surface materials are selected to reflect realistic acoustic conditions. All materials are frequency-dependent and characterized by complex surface impedances, employed in the wave-based and pressure-based ISM calculations, with assignments consistent with their physical counterparts (e.g., glass for windows, wood or plastic for furniture, and gypsum or concrete for walls and ceilings).

Each room contains four randomly placed sources. Most are directive and represent speech sources or loudspeakers with randomized orientations. In larger rooms, one speech source is replaced by an omnidirectional source to increase variability. Each room includes 20–30 receivers, randomly distributed and positioned at least \SI{1}{\meter} from any source and \SI{0.5}{\meter} from any surface.

A hybrid simulation approach is used. The wave-based solver operates up to a crossover frequency between 1 and \SI{2}{\kilo\Hz}, depending on room size, while higher frequencies up to \SI{12}{\kilo\Hz} are simulated using a geometric acoustics (GA) solver that combines the image-source method (order 8, vs 10 in ISM dataset) with a ray-radiosity approach.

Compared with ISM, the Hybrid dataset differs in geometry, furnishing, frequency-dependent materials, and simulation method. These factors jointly increase spectral and temporal diversity; therefore, the study evaluates overall simulation realism rather than any individual modelling component.

\subsection{Training configurations}

\begin{table}[]
\centering
\caption{DeepFilterNet3 training configuration}
\label{tab:dfn_config}
\begin{tabular}{l c}
\hline
\textbf{Parameter} & \textbf{Value} \\
\hline
Network version & $0.5.7pre$ \\
Network type & deepfilternet3 \\
$DRR$ & 0.3 \\
$RT60_\mathrm{target}$ & \SI{0.05}{\second} \\
Late reverberation offset & \SI{5}{\milli\second} \\
$p_{\mathrm{reverb}}$ & 1.00 \\
Epochs & 30 / 120 \\
Speech dataset & DNS4 english \\
Noises & DNS4 \\
Training sampling rate &  \SI{48000}{\hertz} \\
\hline
\end{tabular}
\end{table}


DeepFilterNet3 \cite{schroeter2023deepfilternet3} combines ERB-based gain estimation with multi-frame complex filtering for low-latency denoising and dereverberation. Training mixtures are generated on-the-fly. The parameter $p_{\mathrm{reverb}}$ controls whether speech and noise are convolved with a randomly selected RIR before mixing. Following the original implementation, the same RIR is applied to both components. We retain this pipeline for comparability with prior DeepFilterNet3 studies and to isolate the effect of the RIR dataset.

DeepFilterNet3 also applies stochastic decay augmentation with probability $p_{\mathrm{decay}}$, inherited from $p_{\mathrm{reverb}}$. When enabled, the late RIR tail is modified to an RT$_{60}$ between 0.2 and \SI{1.0}{\second}. Because this may alter the physically simulated decay of Hybrid RIRs, we compare the default setting ($p_{\mathrm{reverb}}=p_{\mathrm{decay}}=1.0$, Cases~1--2) with decay augmentation disabled (Cases~3--4).

The configurations in Table~\ref{tab:exp_matrix} vary training-data size, epoch count, and decay augmentation to test whether the dataset comparison is robust across training conditions.

Because DeepFilterNet3 estimates the direct-path arrival for delay compensation, we retain only Hybrid RIRs with source--receiver line of sight, yielding \num{29144} RIRs and reducing errors in direct-path estimation.

\begin{table}[]
\centering
\caption{Overview of experimental configurations. Speech samples extracted from DNS4 English dataset with the addition of PTDB-TUG. $^\dagger$ RT decay randomization from
DFN3 disabled}
\label{tab:exp_matrix}
\begin{tabular}{l c c c}
\hline
\textbf{Speech} & \textbf{Config} & \textbf{RIR type} & \textbf{Case} \\
\hline
Small 30 ep. & Default        & Hybrid & 1 \\
  &          & ISM    & 2 \\
  \midrule
Small $^\dagger$ 30 ep. & No decay aug.  & Hybrid & 3 \\
  &    & ISM    & 4 \\
  \midrule
Full 30 ep.         & Default        & Hybrid & 5 \\
         &         & ISM    & 6 \\
  \midrule
Full 120 ep.         & Default        & Hybrid & 7 \\
        &         & ISM    & 8 \\
\hline
\end{tabular}
\end{table}

\subsection{Training input}

The reduced speech set comprised approximately 30~GB of LibriVox reading speech, CREMA-D \cite{cao2014crema}, PTDB-TUG, and VCTK, using a sampling factor of 0.2 to preserve the original DeepFilterNet3 balance. Noise was drawn from the DNS4 AudioSet and Freesound data. DNS4 audio was sampled at \SI{48}{\kilo\hertz}; Hybrid RIRs were generated at \SI{32}{\kilo\hertz} and resampled by DeepFilterNet3.

Hybrid RIRs are generated at 32~kHz because material absorption data are available only through the 8~kHz octave band, corresponding to about 12~kHz usable bandwidth. Higher sampling rates would therefore add no physically grounded content in the current solver. Preliminary tests of native 32~kHz training and alternative resampling were less stable, so we retained the original 48~kHz DeepFilterNet3 pipeline. Evaluation uses measured RIRs, and we do not expect this processing choice to bias the dataset comparison.

Speech and RIRs were split 70/15/15 at speaker and room level, respectively. Besides the reduced speech set, we trained on the full DNS4 English set for 30 and 120 epochs (Cases~5--8), with the learning-rate schedule adjusted accordingly.

\section{Evaluation}
\label{sec:evaluation}

Evaluation mixtures used unseen VCTK speech, unseen measured RIRs, and unseen AudioSet and Freesound noise clips excluded from training. The set comprised 827 VCTK utterances \cite{veaux2017vctk} and 284 measured RIRs from ACE \cite{eaton2016ace} and MIT \cite{traer2016statistics}. Speech and noise were reverberated before mixing, with SNR uniformly sampled from \SI{0}{\decibel} to \SI{40}{\decibel}. The evaluation code\footnote{Code and configurations: \url{https://github.com/TrebleTechnologies/iwaenc2026milo}} follows the fully reverberant protocol of Gusó et al.~\cite{guso2025mb}.



To evaluate enhancement performance, we consider objective metrics that capture complementary aspects of speech quality, intelligibility, and distortion, namely Perceptual Evaluation of Speech Quality (PESQ), scale-invariant signal-to-distortion ratio improvement (SI-SDRi), Short-Time Objective Intelligibility (STOI), and Speech-to-Reverberation Modulation Energy Ratio (SRMR). As an additional step, the enhanced files were transcribed using NVIDIA NeMo 1.23.0 \cite{kuchaiev2019nemo, nvidia_nemo_1_23_0}. The ground truth was loaded from the transcripts of the VCTK test sets. The clean and noisy files scored respectively 0.0224 and 0.1671 WER.

\section{Results}
\label{sec:results}

\begin{table}[]
\caption{Objective evaluation metrics (median) for DFN3 trained with ISM and Hybrid RIR datasets across different training regimes.$^\dagger$ RT decay randomization from DFN3 disabled.}
\centering
\begin{tabular}{llccc}
\toprule
\textbf{Setting} & \textbf{Metric} & \textbf{ISM} & \textbf{Hybrid} & $\boldsymbol{\Delta}$ \\
\midrule

\multirow{4}{*}{Small 30 ep.}
 & PESQ   & 2.17 & \textbf{2.35} & \textbf{+0.18} \\
 & SISDRi & 1.20 & \textbf{1.41} & \textbf{+0.21} \\
 & STOI   & 0.700 & \textbf{0.714} & \textbf{+0.014} \\
 & SRMR   & 9.80 & \textbf{9.97} & \textbf{+0.18} \\

\midrule

\multirow{4}{*}{Small $^\dagger$ 30 ep.}
 & PESQ   & 2.18 & \textbf{2.39} & \textbf{+0.20} \\
 & SISDRi & 1.30 & \textbf{1.35} & \textbf{+0.05} \\
 & STOI   & 0.696 & \textbf{0.712} & \textbf{+0.015} \\
 & SRMR   & 10.33 & \textbf{10.42} & \textbf{+0.09} \\

\midrule

\multirow{4}{*}{Full 30 ep.}
 & PESQ   & 2.24 & \textbf{2.41} & \textbf{+0.17} \\
 & SISDRi & 1.34 & \textbf{1.46} & \textbf{+0.12} \\
 & STOI   & 0.711 & \textbf{0.724} & \textbf{+0.013} \\
 & SRMR   & 9.86 & \textbf{10.05} & \textbf{+0.19} \\

\midrule

\multirow{4}{*}{Full 120 ep.}
 & PESQ   & 2.27 & \textbf{2.43} & \textbf{+0.16} \\
 & SISDRi & 1.35 & \textbf{1.55} & \textbf{+0.21} \\
 & STOI   & 0.717 & \textbf{0.729} & \textbf{+0.012} \\
 & SRMR   & 9.78 & \textbf{10.33} & \textbf{+0.55} \\

\bottomrule
\end{tabular}

\label{tab:RTON_all}
\end{table}

\begin{table}[]
\caption{Overall paired improvement of Hybrid over ISM across all configurations. Values are mean differences with 95\% bootstrap confidence intervals paired at utterance level.}
\centering
\begin{tabular}{lcc}
\toprule
\textbf{Metric} & $\boldsymbol{\Delta}$ (Hybrid--ISM) & \textbf{95\% CI} \\
\midrule
PESQ   & \textbf{+0.166} & \textbf{[+0.149, +0.184]} \\
SISDRi & \textbf{+0.110} & \textbf{[+0.040, +0.182]} \\
STOI   & \textbf{+0.013} & \textbf{[+0.010, +0.015]} \\
SRMR   & \textbf{+0.270} & \textbf{[+0.213, +0.328]} \\
\bottomrule
\end{tabular}

\label{tab:summary}
\end{table}

\begin{table}[]
\caption{ASR performance (WER) using NeMo on enhanced speech. $\Delta$ vs noisy denotes the absolute WER reduction with respect to the noisy baseline (0.1671), while Rel. imp. (\%) denotes the corresponding relative improvement. 
Training with the Hybrid (Treble) dataset consistently yields lower WER than training with the ISM (DNS4) dataset across all evaluated configurations. $^\dagger$ RT decay randomization from DFN3 disabled.}
\centering
\begin{tabular}{lccc}
\toprule
\textbf{Model} & \textbf{WER} $\downarrow$ & $\boldsymbol{\Delta}$ vs noisy & \textbf{Rel. imp. (\%)} \\
\midrule
\textbf{Clean reference} & \textbf{0.0224} & +0.1447 & -- \\
\midrule
\multicolumn{4}{c}{\textit{Hybrid (Treble) training}} \\
\midrule
Full 30 ep.      & \textbf{0.1274} & \textbf{+0.0397} & \textbf{23.8\%} \\
Full 120 ep.     & 0.1290 & +0.0381 & 22.8\% \\
Small $^\dagger$ 30 ep.   & 0.1399 & +0.0272 & 16.3\% \\
Small 30 ep.         & 0.1470 & +0.0201 & 12.0\% \\
\midrule
\multicolumn{4}{c}{\textit{ISM (DNS4) training}} \\
\midrule
Full 30 ep.      & 0.1452 & +0.0219 & 13.1\% \\
Full 120 ep.     & 0.1492 & +0.0179 & 10.7\% \\
Small $^\dagger$ 30 ep.  & 0.1616 & +0.0055 & 3.3\% \\
Small 30 ep.        & 0.1663 & +0.0008 & 0.5\% \\
\midrule
\textbf{Noisy baseline}  & 0.1671 & +0.0000 & 0.0\% \\
\bottomrule
\end{tabular}

\label{tab:wer_main}
\end{table}

\subsection{Objective Evaluation} 



Table~\ref{tab:RTON_all} shows that training with the Hybrid RIR dataset consistently yields modest but positive improvements over the ISM dataset across all objective metrics and training configurations.
In the small 30-epoch setting, Hybrid improves PESQ, SI-SDRi, STOI, and SRMR by +0.18, +0.21, +0.014, and +0.18, respectively. Similar gains are observed when RT decay randomization is disabled, although the SI-SDRi improvement is smaller (+0.05), suggesting that reverberation decay augmentation mainly benefits distortion reduction. The same trend holds for the full-data models: Hybrid improves all metrics at both 30 and 120 epochs, with the full 120-epoch setting yielding the strongest overall results. 

Averaged across all configurations, Table~\ref{tab:summary} confirms a systematic advantage of Hybrid, with positive mean gains and 95\% bootstrap confidence intervals that are strictly positive for all four metrics. These objective improvements are accompanied by larger gains in downstream ASR. 

The confidence intervals were computed from utterance-level paired differences pooled across all configurations and therefore represent the average paired improvement of the Hybrid dataset over ISM rather than variability within individual training settings.

Table~\ref{tab:wer_main} shows that Hybrid achieves lower WER than ISM in every matched configuration, with relative improvements over the noisy baseline ranging from 12.0\% to 23.8\%, compared to 0.5\% to 13.1\% for ISM. The largest gains are obtained for the full-data models, but the same pattern holds in the small-data setting. 


Although the improvements in conventional speech enhancement metrics are relatively modest, they are remarkably consistent across all evaluated training configurations. More importantly, these gains are accompanied by substantially larger improvements in downstream ASR performance. This suggests that training with the higher-fidelity RIR dataset better preserves speech characteristics relevant for automatic recognition than is reflected by conventional perceptual metrics alone.

We emphasize, however, that the present experiments compare complete RIR generation pipelines rather than individual acoustic modelling components. Consequently, the observed improvements should be interpreted as evidence that increasing the overall realism of the synthetic training data improves generalization to unseen measured environments, rather than as evidence for the contribution of any single factor, such as hybrid wave-based simulation, frequency-dependent material modelling, or room geometry.


\subsection{Limitations}


The objective of this work is to compare two complete RIR generation pipelines representative of commonly used and higher-fidelity room acoustic simulation. Consequently, several realism factors differ simultaneously between the datasets, including room geometry, furnishing, frequency-dependent material properties, and the underlying acoustic simulation method. While this holistic comparison reflects practical dataset generation workflows, it does not permit attribution of the observed performance gains to any individual modelling component.

Future work will therefore investigate controlled ablations using datasets in which specific acoustic factors, such as frequency-dependent absorption, room geometry, or wave-based low-frequency simulation, can be varied independently. Additional perceptual evaluation, including listening tests and learned quality metrics such as DNSMOS or UTMOS, would also provide complementary insight into the perceptual significance of the observed improvements.

\section{Conclusion}
\label{sec:Conclusion}

This work evaluated how the overall realism of synthetic room impulse response (RIR) datasets influences the training of DeepFilterNet3 for single-channel speech enhancement. Across all evaluated training configurations, models trained with the higher-fidelity synthetic RIR dataset consistently achieved modest improvements in objective speech enhancement metrics together with substantially larger improvements in downstream ASR performance on unseen measured environments.

Rather than attributing these gains to any individual acoustic modelling component, our results indicate that increasing the realism of the synthetic training data leads to improved model generalization. These findings complement previous studies that isolated specific simulation factors by showing that, from a practical dataset generation perspective, adopting a more realistic room acoustic simulation pipeline can provide consistent benefits for both speech enhancement and downstream recognition tasks. Additionally, while longer training produced modest improvements in objective enhancement metrics, these gains did not consistently translate into lower ASR word error rates, suggesting that conventional enhancement metrics and downstream recognition performance do not necessarily evolve in parallel.

Future work will investigate which aspects of simulation realism contribute most strongly to these improvements through controlled ablation studies, including room geometry, material modelling, and frequency-dependent acoustic simulation, as well as broader perceptual evaluation using listening tests and learned quality metrics.






\balance

{\footnotesize
\bibliographystyle{IEEEbib}
\bibliography{refs}

@STRING{JASA = {J. Acoust. Soc. Am.}}

@STRING{SoundVib = {J. Sound Vib.}}

@STRING{ICASSP = {Proc. IEEE Int. Conf. Acoust. Speech Signal Process.~(ICASSP)}}

@STRING{ApplAcoust = {Appl. Acoust.}}

@STRING{ACMTOG = {ACM Trans. Graph.}}

@STRING{WASPAA = {Proc. IEEE Workshop Appl. Signal Process. Audio Acoust.~(\mbox{WASPAA})}}

@STRING{ISRA = {Proc. Int. Symp. Room Acoust. (ISRA)}}

@STRING{SIGGRAPH = {Proc. ACM SIGGRAPH Conf.}}

@STRING{PNAS = {Proc. Natl. Acad. Sci.~(PNAS)}}

@STRING{Interspeech = {Proc. Interspeech}}

@article{kuchaiev2019nemo,
  title={NeMo: a toolkit for building AI applications using Neural Modules},
  author={Kuchaiev, Oleksii and Li, Jason and Nguyen, Huyen and Hrinchuk, Oleksii and Leary, Ryan and Ginsburg, Boris and Kriman, Samuel and Beliaev, Stanislav and Lavrukhin, Vitaly and Cook, Jack and Castonguay, Patrice and Popova, Mariya and Huang, Jocelyn and Cohen, Jonathan M.},
  journal={arXiv:1909.09577},
  year={2019},
  note={NeurIPS 2019 Workshop on Machine Learning Systems}
}

@software{nvidia_nemo_1_23_0,
  author    = {{NVIDIA Corporation}},
  title     = {{NeMo}: A toolkit for {C}onversational {AI}},
  version   = {1.23.0},
  year      = {2024},
  url       = {https://github.com/NVIDIA-NeMo/NeMo/releases/tag/v1.23.0},
  note      = {Software package, version 1.23.0, 2024.}
}

@inproceedings{veaux2017vctk,
  title     = {CSTR VCTK Corpus: English Multi-speaker Corpus for CSTR Voice Cloning Toolkit (Version 0.92)},
  author    = {Veaux, Christophe and Yamagishi, Junichi and MacDonald, Kirsten},
  booktitle = {University of Edinburgh, CSTR},
  year      = {2017}
}

@article{traer2016statistics,
  title   = {Statistics of Natural Reverberation Enable Perceptual Separation of Sound and Space},
  author  = {Traer, James and McDermott, Josh H.},
  journal = {Proc. Natl. Acad. Sci. (PNAS)},
  volume  = {113},
  number  = {48},
  pages   = {E7854--E7863},
  year    = {2016}
}

@article{eaton2016ace,
  title   = {Estimation of Room Acoustic Parameters: The ACE Challenge},
  author  = {Eaton, James and Gaubitch, Nikolay D. and Moore, Alastair H. and Naylor, Patrick A.},
  journal = {IEEE/ACM Trans. Audio Speech Lang. Process.},
  volume  = {24},
  number  = {10},
  pages   = {1681--1693},
  year    = {2016}
}

@article{cao2014crema,
  title={CREMA-D: Crowd-sourced Emotional Multimodal Actors Dataset},
  author={Cao, Houwei and Cooper, David G. and Keutmann, Michael K. and Gur, Ruben C. and Nenkova, Ani and Verma, Ragini},
  journal={IEEE Transactions on Affective Computing},
  year={2014}
}

@article{rosenbaum2024deepfiltering,
  title   = {Deep Learning Framework for Efficient Real-Time Speech Enhancement and Dereverberation},
  author  = {Rosenbaum, Tomer and Winebrand, Emil and Cohen, Omer and Cohen, Israel},
  journal = {Sensors},
  volume  = {24},
  number  = {3},
  pages   = {630},
  year    = {2024},
  publisher = {MDPI},
  doi     = {10.3390/s24030630}
}

@article{wu2025egonoise,
  title={Egonoise Resilient Source Localization and Speech Enhancement for Drones Using a Hybrid Model and Learning-Based Approach},
  author={Wu, Yihsuan and Chiu, Yukai and Anthony, Michael and Bai, Mingsian R.},
  journal={arXiv:2508.06310},
  year={2025}
}

@article{wang2018supervised,
  title={Supervised Speech Separation Based on Deep Learning: An Overview},
  author={Wang, DeLiang and Chen, Jitong},
  journal={IEEE/ACM Trans. Audio Speech Lang. Process.},
  volume={26},
  number={10},
  pages={1702--1726},
  year={2018},
  publisher={IEEE}
}

@inproceedings{westhausen2020dtln,
  title={Dual-Signal Transformation LSTM Network for Real-Time Noise Suppression},
  author={Westhausen, Nils L. and Meyer, Bernd T.},
  booktitle={Proc. Interspeech},
  year={2020},
  pages={2477--2481}
}

@inproceedings{tan2018convolutional,
  title={A convolutional recurrent neural network for real-time speech enhancement.},
  author={Tan, Ke and Wang, DeLiang},
  booktitle={Interspeech},
  volume={2018},
  pages={3229--3233},
  year={2018}
}

@inproceedings{schroeter2023deepfilternet3,
  title = {{DeepFilterNet}: Perceptually Motivated Real-Time Speech Enhancement},
  author = {Schröter, Hendrik and Rosenkranz, Tobias and Escalante-B., Alberto N. and Maier, Andreas},
  booktitle={INTERSPEECH},
  year = {2023},
}

@inproceedings{guso2025mb,
  title={{MB-RIRs}: a Synthetic Room Impulse Response Dataset with Frequency-Dependent Absorption Coefficients},
  author={Gus{\'o}, Enric and Luberadzka, Joanna and Sayin, Umut and Serra, Xavier},
  booktitle=WASPAA,
  pages={1--5},
  year={2025},
  address={},
}

@inproceedings{schroter2022deepfilternet,
  title={{DeepFilterNet}: A low complexity speech enhancement framework for full-band audio based on deep filtering},
  author={Schröter, Hendrik and Escalante-B., Alberto N and Rosenkranz, Tobias and Maier, Andreas},
  booktitle=ICASSP,
  pages={7407--7411},
  year={2022},
  address = {},
}

@inproceedings{arakawa2024quantifying,
  title={Quantifying the effect of simulator-based data augmentation for speech recognition on augmented reality glasses},
  author={Arakawa, Riku and Parvaix, Mathieu and Lai, Chiong and Erdogan, Hakan and Olwal, Alex},
  booktitle=ICASSP,
  pages={726--730},
  year={2024},
  address={}
}

@inproceedings{bezzam2020study,
  title={A study on more realistic room simulation for far-field keyword spotting},
  author={Bezzam, Eric and Scheibler, Robin and Cadoux, Cyril and Gisselbrecht, Thibault},
  booktitle={Proc. Asia-Pacific Signal Information Process. Assoc. Annual Summit Conf. (APSIPA ASC)},
  pages={674--680},
  year={2020},
  address={},
}

@article{Schissler2014HighOrderDiffractionAndDiffuseReflectionsForInteractiveSoundPropagation,
	author = {Schissler, Carl and Mehra, Ravish and Manocha, Dinesh},
	title = {{High-order diffraction and diffuse reflections for interactive sound propagation in large environments}},
	issn = {0730-0301},
	doi = {10.1145/2601097.2601216},
	number = {4, article no. 39},
	volume = {33},
	journal = ACMTOG,
	year = {2014}
}

@article{Borish1984ExtensionOfTheImageModelToArbitraryPolyhedra,
	author = {Borish, Jeffrey},
	title = {{Extension of the image model to arbitrary polyhedra}},
	doi = {10.1121/1.390983},
	pages = {1827--1836},
	number = {6},
	volume = {75},
	journal = JASA,
	year = {1984}
}

@article{Allen1979ImageMethodForEfficientlySimulatingSmallRoomAcoustics,
	author = {Allen, Jont B. and Berkley, David A.},
	title = {{Image method for efficiently simulating small‐room acoustics}},
	doi = {10.1121/1.382599},
	pages = {943--950},
	number = {4},
	volume = {65},
	journal = JASA,
	year = {1979}
}

@article{Krokstad1968CalculatingTheAcousticalRoomResponseByTheUseOfRayTracing,
	author = {Krokstad, A. and Strøm, S. and Sørsdal, S.},
	title = {{Calculating the acoustical room response by the use of a ray tracing technique}},
	doi = {10.1016/0022-460x(68)90198-3},
	pages = {118--125},
	number = {1},
	volume = {8},
	journal = SoundVib,
	year = {1968}
}

@article{Krokstad1983Fifteenyearsexperiencewithcomputerizedraytracing, 
  year     = {1983}, 
  title    = {Fifteen years' experience with computerized ray tracing}, 
  author   = {Krokstad, A and Strøm, S and Sørsdal, S}, 
  journal  = ApplAcoust, 
  doi      = {10.1016/0003-682x(83)90021-x}, 
  pages    = {291--312}, 
  number   = {4}, 
  volume   = {16}
}

@article{Brinkmann2019RoundRobinOnRoomAcousticalSimulationAndAuralization,
	author = {Brinkmann, Fabian and Aspöck, Lukas and Ackermann, David and Lepa, Steffen and Vorländer, Michael and Weinzierl, Stefan},
	title = {{A round robin on room acoustical simulation and auralization}},
	doi = {10.1121/1.5096178},
	pages = {2746--2760},
	number = {4},
	volume = {145},
	journal = JASA,
	year = {2019}
}

@article{BottelDooren1995FDTDSimulationRoomAcoustic, 
	year = {1995}, 
	title = {{Finite-difference time-domain simulation of low-frequency room acoustic problems}}, 
	author = {Botteldooren, D}, 
	journal = JASA, 
	doi = {10.1121/1.413817}, 
	pages = {3302--3308}, 
	number = {6}, 
	volume = {98}, 
}

@article{Prinn2023ReviewFEMRoomAcoustics, 
	year = {2023}, 
	title = {A Review of Finite Element Methods for Room Acoustics}, 
	author = {Prinn, Albert G.}, 
	journal = {Acoustics}, 
	doi = {10.3390/acoustics5020022}, 
	pages = {367--395}, 
	number = {2}, 
	volume = {5}, 
}

@article{Pind2019SpectralElementRoomAcousticSimulation, 
	year = {2019}, 
	title = {{Time domain room acoustic simulations using the spectral element method}}, 
	author = {Pind, Finnur and Engsig-Karup, Allan P. and Jeong, Cheol-Ho and Hesthaven, Jan S. and Mejling, Mikael S. and Strømann-Andersen, Jakob}, 
	journal = JASA, 
	doi = {10.1121/1.5109396}, 
	pages = {3299--3310}, 
	number = {6}, 
	volume = {145}, 
}

@article{Wang2019DiscontinuousGalerkinRoomAcoustics, 
	year = {2019}, 
	title = {{Room acoustics modelling in the time-domain with the nodal discontinuous Galerkin method}}, 
	author = {Wang, Huiqing and Sihar, Indra and Pagán Muñoz, Raúl and Hornikx, Maarten}, 
	journal = JASA, 
	doi = {10.1121/1.5096154}, 
	pages = {2650--2663}, 
	number = {4}, 
	volume = {145}, 
}

@article{Pind2020TimeDomainSimExtendedReactingPorousDGFEM, 
year = {2020}, 
title = {{Time-domain room acoustic simulations with extended-reacting porous absorbers using the discontinuous Galerkin method}}, 
author = {Pind, Finnur and Jeong, Cheol-Ho and Engsig-Karup, Allan P. and Hesthaven, Jan S. and Strømann-Andersen, Jakob}, 
journal = JASA, 
doi = {10.1121/10.0002448}, 
pages = {2851--2863}, 
number = {5}, 
volume = {148}
}

@article{Melander2024MassivelyParallelGalerkinRoomAcoustics, 
	year = {2024}, 
	title = {{Massively parallel nodal discontinous Galerkin finite element method simulator for room acoustics}}, 
	author = {Melander, Anders and Strøm, Emil and Pind, Finnur and Engsig-Karup, Allan P and Jeong, Cheol-Ho and \mbox{Warburton}, Tim and Chalmers, Noel and Hesthaven, Jan S}, 
	journal = {Int. J. High Perform. Comput. Appl.}, 
	doi = {10.1177/10943420231208948}, 
	pages = {154--174}, 
	number = {3}, 
	volume = {38}, 
}

@article{Savioja2015OverviewGA, 
	year = {2015}, 
	title = {{Overview of geometrical room acoustic modeling techniques}}, 
	author = {Savioja, Lauri and Svensson, U Peter}, 
	journal = JASA, 
	doi = {10.1121/1.4926438}, 
	pages = {708--730}, 
	number = {2}, 
	volume = {138}, 
}

@phdthesis{Hamilton2016FiniteDifferenceVolumeWaveBasedRoomAcousticsThesis, 
year = {2016}, 
title = {Finite Difference and Finite Volume Methods for Wave-based Modelling of Room Acoustics}, 
author = {Hamilton, Brian}, 
institution = {University of Edinburgh},
school = {University of Edinburgh}
}

@article{Vorlaender2013ComputerSimulationsRoomAcoustics, 
	year = {2013}, 
	title = {{Computer simulations in room acoustics: Concepts and uncertainties}}, 
	author = {Vorländer, Michael}, 
	journal = JASA, 
	doi = {10.1121/1.4788978}, 
	pages = {1203--1213}, 
	number = {3}, 
	volume = {133}, 
}

@article{Torres2001ComputationEdgeDiffractionRoomAcousticSimulations, 
	year = {2001}, 
	title = {{Computation of edge diffraction for more accurate room acoustics auralization}}, 
	author = {Torres, Rendell R. and Svensson, U. Peter and Kleiner, Mendel}, 
	journal = JASA, 
	doi = {10.1121/1.1340647}, 
	pages = {600--610}, 
	number = {2}, 
	volume = {109}
}

@article{aralikatti2022synthetic,
  title={Synthetic wave-geometric impulse responses for improved speech dereverberation},
  author={Aralikatti, Rohith and Tang, Zhenyu and Manocha, Dinesh},
  journal={arXiv:2212.05360},
  year={2022}
}

@inproceedings{Tang2022GWALargeDatasetForAudioProcessing, 
	year = {article no. 36, 2022}, 
	author = {Tang, Zhenyu and Aralikatti, Rohith and Ratnarajah, Anton Jeran and Manocha, Dinesh}, 
	title = {{GWA}: A Large High-Quality Acoustic Dataset for Audio Processing}, 
	booktitle = SIGGRAPH, 
	doi = {10.1145/3528233.3530731}, 
	address = {}, 
}

@INPROCEEDINGS{dubey2022icassp2022deepnoise,
  author={Dubey, Harishchandra and Gopal, Vishak and Cutler, Ross and Aazami, Ashkan and Matusevych, Sergiy and Braun, Sebastian and Eskimez, Sefik Emre and Thakker, Manthan and Yoshioka, Takuya and Gamper, Hannes and Aichner, Robert},
  booktitle={Proc. IEEE Int. Conf. Acoust., Speech Signal Process. (ICASSP), 2022}, 
  title={Icassp 2022 Deep Noise Suppression Challenge}, 
  year={2022},
  volume={},
  number={},
  pages={9271-9275},
  doi={10.1109/ICASSP43922.2022.9747230}}

@inproceedings{Ko2017DataAugmentationReverberantSpeechForSpeechRecognition, 
	year = {2017}, 
	title = {A Study on Data Augmentation of Reverberant Speech for Robust Speech Recognition}, 
	author = {Ko, Tom and Peddinti, Vijayaditya and Povey, Daniel and Seltzer, Michael L. and Khudanpur, Sanjeev}, 
	booktitle = ICASSP, 
	address = {},
	doi = {10.1109/icassp.2017.7953152}, 
	pages = {5220--5224}, 
}

@inproceedings{Siltanen2010RaysOrWavesStrenghtsAndWeaknesses, 
year = {2010}, 
author = {Siltanen, Samuel and Lokki, Tapio and Savioja, Lauri}, 
title = {Rays or Waves? Understanding the Strengths and Weaknesses of Computational Room Acoustics Modeling Techniques}, 
booktitle = ISRA, 
address = {}, 
}

@inproceedings{Kim2017GenerationLargeScaleSimulatedUtterancesFarFieldASR, 
  year    = {2017}, 
  title   = {Generation of Large-Scale Simulated Utterances in Virtual Rooms to Train Deep-Neural Networks for Far-Field Speech Recognition in {Google Home}}, 
  author  = {Kim, Chanwoo and Misra, Ananya and Chin, Kean and Hughes, Thad and Narayanan, Arun and Sainath, Tara and Bacchiani, Michiel}, 
  booktitle = Interspeech, 
  doi     = {10.21437/interspeech.2017-1510}, 
  pages   = {379--383},
  address = {},
}

@inproceedings{Maciejewski2020WHAMR, 
  year     = {2020}, 
  title    = {{WHAMR}!: Noisy and Reverberant Single-Channel Speech Separation}, 
  author   = {Maciejewski, Matthew and Wichern, Gordon and {McQuinn}, Emmett and Le Roux, Jonathan}, 
  booktitle  = ICASSP, 
  doi      = {10.1109/icassp40776.2020.9053327}, 
  pages    = {696--700}, 
  address = {Online conference},
}

@inproceedings{Srivastava2023HowToVirtuallyTrainYourSpeakerLocalizer, 
  year     = {2023}, 
  title    = {How to (Virtually) Train Your Speaker Localizer}, 
  author   = {Srivastava, Prerak and Deleforge, Antoine and Politis, Archontis and Vincent, Emmanuel}, 
  booktitle  = Interspeech, 
  doi      = {10.21437/interspeech.2023-1065},  
  pages    = {1204--1208},
  address = {},
}
}

\end{document}